\documentclass[a4paper,11pt]{article}
\usepackage{subfig}

\pdfoutput=1 % if your are submitting a pdflatex (i.e. if you have
\usepackage{ gensymb }
\usepackage{jinstpub} % for details on the use of the package, please
\usepackage{lineno}
\title{Timing performance of radiation hard MALTA monolithic Pixel sensors}

\author[a,1]{G.~Gustavino,\note{Corresponding author.}}
\author[b]{P.~Allport,}
\author[a]{I.~Asensi,}
\author[c]{D.V.~Berlea,}
\author[d]{D.~Bortoletto,}
\author[e]{C.~Buttar,}
\author[a]{F.~Dachs,}
\author[a]{V.~Dao,}
\author[h]{H.~Denizli,}
\author[a,g]{D.~Dobrijevic,}
\author[a]{L.~Flores,}
\author[a]{A.~Gabrielli,}
\author[b]{L.~Gonella,}
\author[f]{V.~González,}
\author[a]{M.~LeBlanc,}
\author[h]{K.~Oyulmaz,}
\author[a]{H.~Pernegger,}
\author[a,i]{F.~Piro,}
\author[a]{P.~Riedler,}
\author[j]{H.~Sandaker,}
\author[a]{C.~Solans,}
\author[a]{W.~Snoeys,}
\author[g]{T.~Suligoj,}
\author[a,j]{M.~van Rijnbach,}
\author[a]{A.~Sharma,}
\author[a,f]{M.~Vázquez Núñez,}
\author[a,k]{J.~Weick,}
\author[c]{S.~Worm,}
\author[k]{A.~Zoubir}

\affiliation[a]{CERN, Geneva, Switzerland}
\affiliation[b]{University of Birmingham, Birmingham, United Kingdom}
\affiliation[c]{DESY, Zeuthen, Germany}
\affiliation[d]{University of Oxford, Oxford, United Kingdom}
\affiliation[e]{University of Glasgow, Glasgow, United Kingdom}
\affiliation[f]{University of Valencia, Valencia, Spain}
\affiliation[g]{University of Zagreb, Zagreb, Croatia}
\affiliation[h]{Bolu Abant Izzet Baysal University, Bolu, Turkey}
\affiliation[i]{EPFL, Lausanne, Schwitzerland}
\affiliation[j]{University of Oslo, Oslo, Norway}
\affiliation[k]{Technische Universit{\"a}t Darmstadt, Darmstadt, Germany}

\emailAdd{giuliano.gustavino@cern.ch}

\abstract{The MALTA family of Depleted Monolithic Active Pixel Sensor (DMAPS) produced in Tower~180~nm CMOS technology targets radiation hard applications for the HL-LHC and beyond. Several process modifications and front-end improvements have resulted in radiation hardness up to $2 \times 10^{15}~1~\text{MeV}~\text{n}_{eq}/\text{cm}^2$ and time resolution below 2~ns, with uniform charge collection efficiency across the Pixel of size $36.4 \times 36.4~\mu\text{m}^2$ with a $3~\mu\text{m}^2$ electrode size. The MALTA2 demonstrator produced in 2021 on high-resistivity epitaxial silicon and on Czochralski substrates implements a new cascoded front-end that reduces the RTS noise and has a higher gain. This contribution shows results from MALTA2 on timing resolution at the nanosecond level from the CERN SPS test-beam campaign of 2021.}

\proceeding{23rd International Workshop on Radiation Imaging Detectors \\
  26-30 June 2022\\
  Riva del Garda, Italy}

\begin{document}
\maketitle
\flushbottom

\section{Introduction}
\label{sec:intro}

High-energy physics experiments require an unprecedented level of precision to measure very rare events in dense environments.
The future harsh pileup conditions foreseen in the Phase-2 at the Large Hadron Collider (LHC) with an average of 200 interactions per bunch crossing will require innovations in detector technologies.
Among the most challenging and important requirements for future tracking systems there are extreme radiation tolerance, high granularity and fast response time.
Then, tracking detectors are requested to cover large surface areas but at the same time to reduce as much as possible the scattering material.
Hybrid silicon pixel detectors, where the ASIC is bump-bonded to the sensors, are the most often adopted and field-tested solution so far in experiments at colliders.
Recently, monolithic pixel sensors have been developed as an interesting alternative.
Such sensors help to minimise the material budget and to reduce construction costs by exploiting industrial CMOS production process of commercial foundries.
The read-out electronics are integrated into the same silicon wafer of the sensor, avoiding the need of custom expensive bump-bonding.
Small electrodes can be designed in order to get low sensor capacitances hence increasing the signal over noise ratio even with a limited thickness of the active layer.
One of the main challenges to face with these detectors is demonstrating the effective radiation hardness up to 100 Mrad in Total Ionizing Dose (TID) and $\geq 1 \times 10^{15}~1~\text{MeV}~\text{n}_{eq}/\text{cm}^2$ in Non-Ionizing Energy Loss (NIEL) in order to be used in the harsh environment of $pp$-collider experiments at LHC and future colliders.
In this context, the MALTA project represents an evolution in MAPS sensors compared to the ALPIDE chip used in the ALICE Inner Tracking System~\cite{ALICE:2013nwm} in Run-3.

\section{The MALTA sensor}
\begin{figure}[htbp]
\centering 
\subfloat[Standard process]{\includegraphics[width=.37\textwidth]{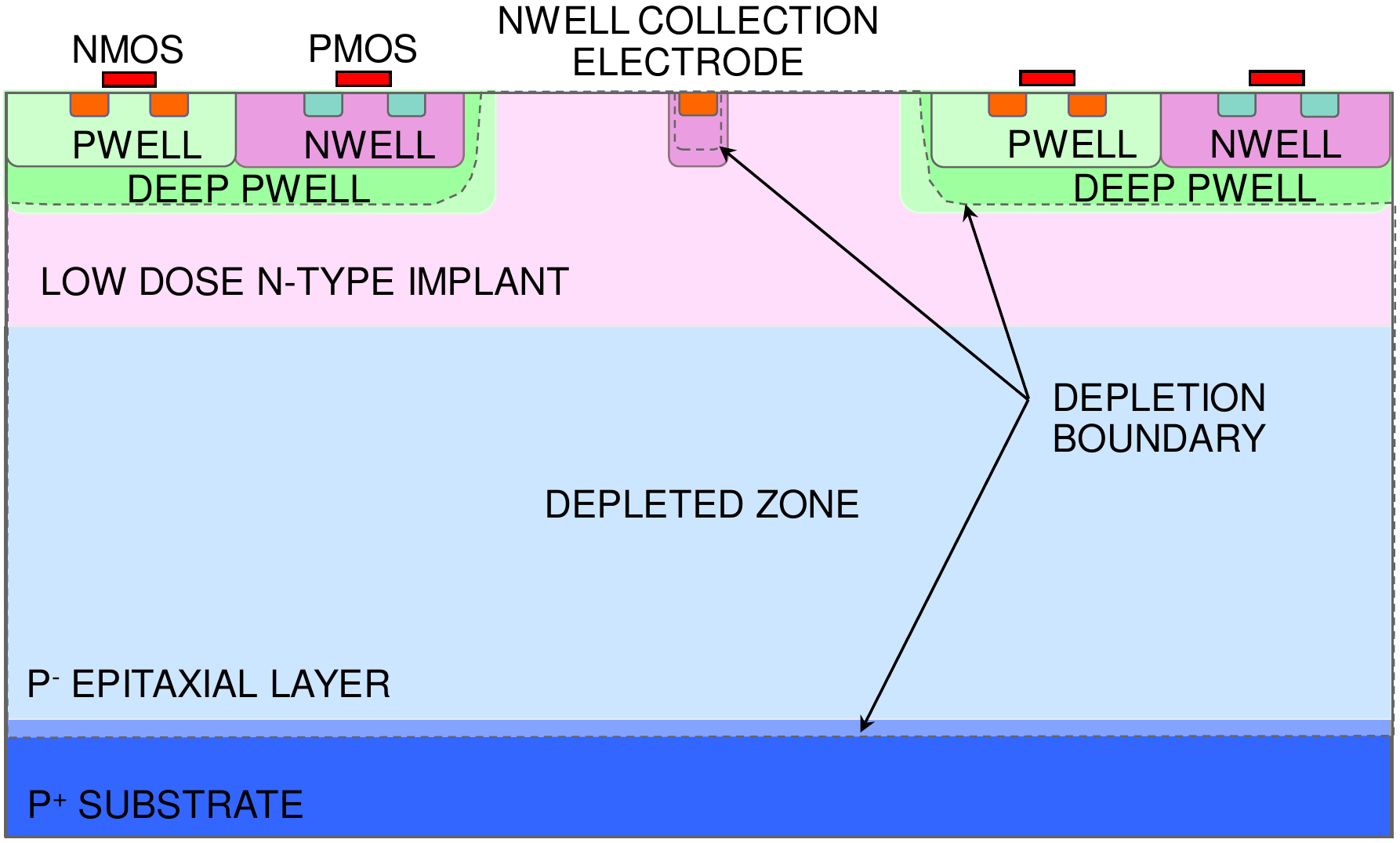}\label{fig:1a}}\\
\subfloat[Gap in the low-dose n-type implant]{\includegraphics[width=.37\textwidth]{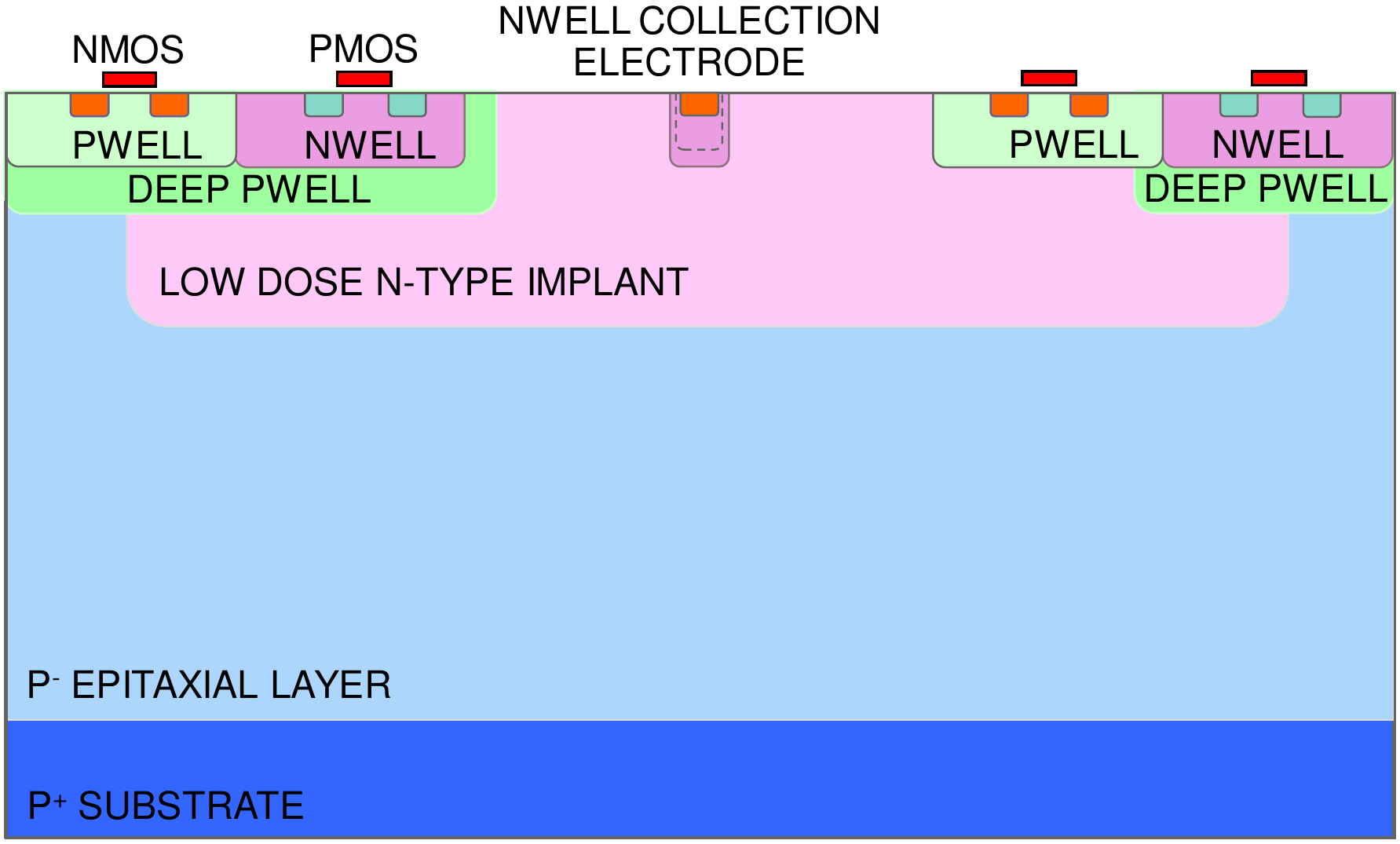}\label{fig:1b}}
\qquad
\subfloat[Extra deep p-well]{\includegraphics[width=.37\textwidth]{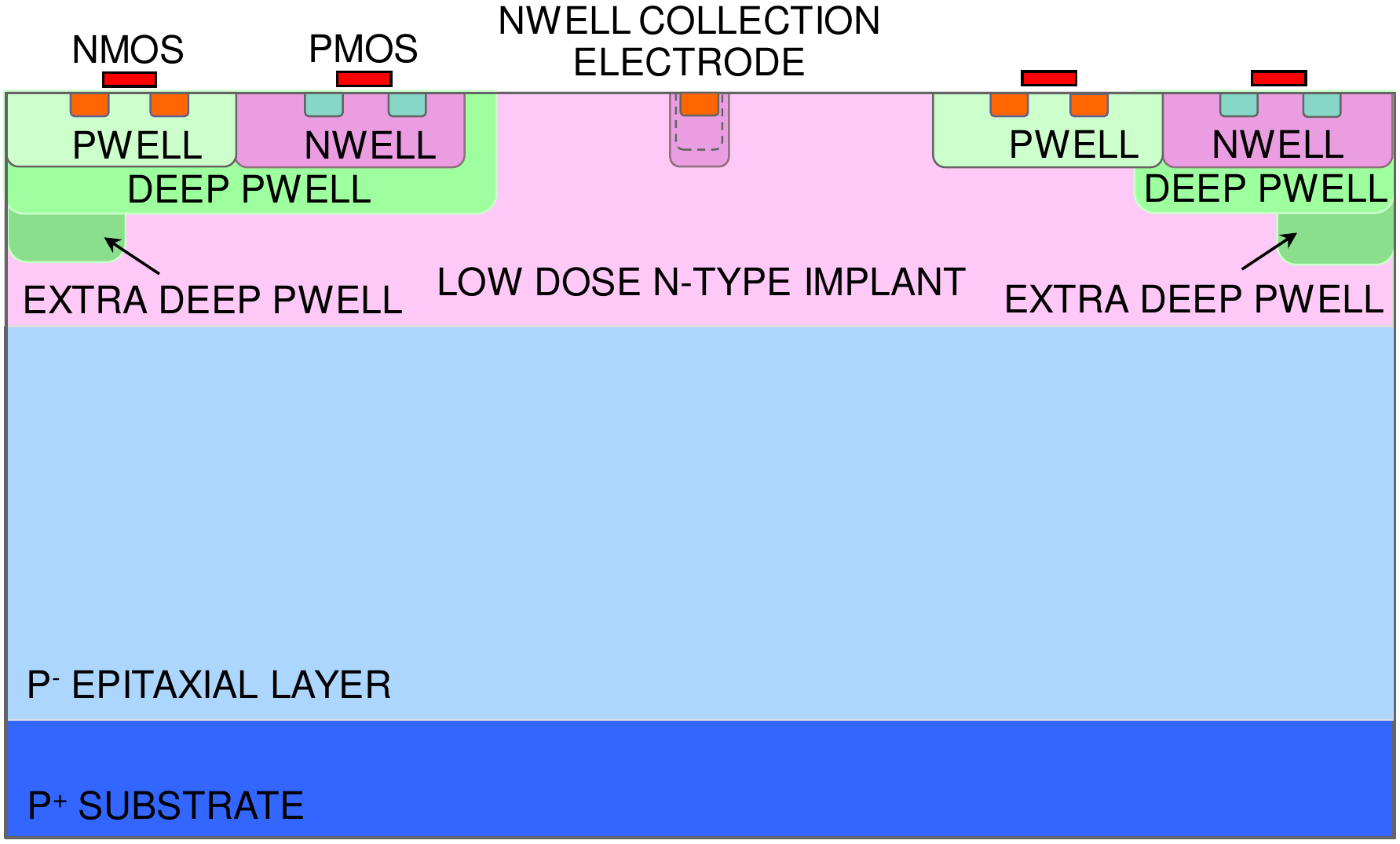}\label{fig:1c}}
\caption{Scheme of the cross section of the of the MALTA sensor in the standard process~(\ref{fig:1a}), with gap in the low-dose n-implant~(\ref{fig:1b}) and with an extra deep p-well~(\ref{fig:1c}).}
\label{fig:1}
\end{figure}

\begin{figure}[htbp]
\centering 
\subfloat[No irradiation]{\includegraphics[width=.4\textwidth]{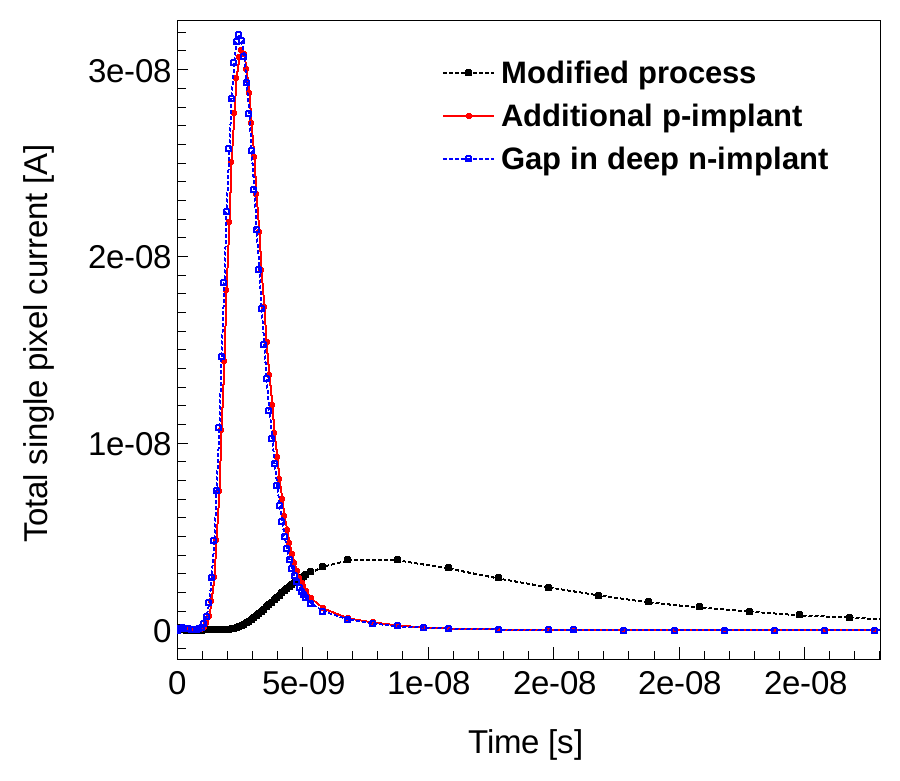}\label{fig:2a}}
\subfloat[$10^{15}~\text{n}_{eq}/\text{cm}^2$]{\includegraphics[width=.4\textwidth]{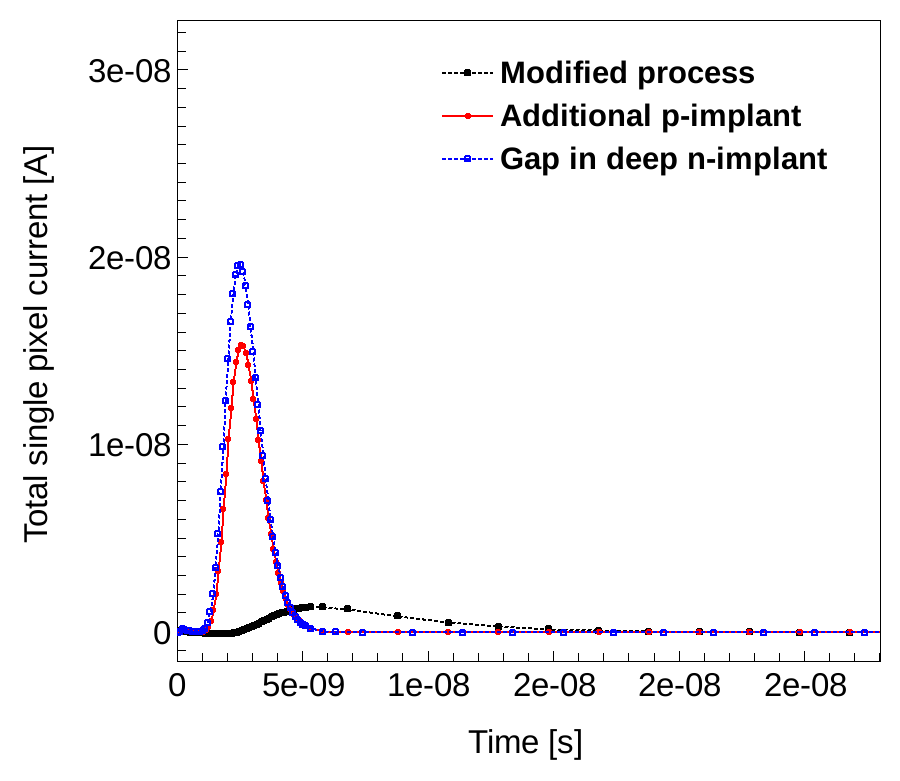}\label{fig:2b}}
\caption{Simulation of the current induced by a MIP incident at the pixel corner as a function of the collection time before~(\ref{fig:2a}) and after irradiation~(\ref{fig:2b})~\cite{Munker:2019vdo}.}
\label{fig:2}
\end{figure}

The MALTA sensor is a Depleted Monolithic Active Pixel Sensor (DMAPS) prototype developed in the Tower Semiconductor 180 nm CMOS imaging process, modified with an additional low dose n-type implant. The pixel size is $36.4 \times 36.4~\mu\text{m}^2$, and thicknesses of 30, 100 and $300~\mu\text{m}$ are tested.
The front-end circuit is optimised to  operate at a threshold of about 200~e$^-$ with a sufficiently fast response for the 25 ns timing requirement of the HL-LHC bunch crossing. 
These features, together with a collection electrode of $3 \times 3~\mu\text{m}^2$, guarantee a minimal capacitance (5 fF) implying a low power consumption ($\sim 1~\mu\text{W / pixel}$).
The sensors are produced with two different substrate kinds: high-resistivity epitaxial layer (Epi) and high-resisitivity ($3-4~k\Omega$) p-type Czochralski (Cz) substrate. The latter is expected to be operational up to higher substrate voltage hence having larger depleted regions and to have a higher radiation resistance and a larger cluster size.
Despite the full depletion of the epitaxial layer, the lateral electric field in the pixel corners is relatively low, resulting in a relatively long collection time and lower efficiency.
Two additional process modifications have been produced in order to address this issue: a gap in the low dose n-type implant (NGAP) or an additional p-type implant at the pixel border (XDPW).
The sensor cross sections in the standard  process, the NGAP and XDPW are shown in figure~\ref{fig:1}.
Figure~\ref{fig:2} shows the TCAD simulation of the current induced by a minimum ionising particle (MIP) traversing the pixel corner as a function of the collection time for the three pixel sensor configurations before and after a NIEL irradiation of $10^{15}~\text{n}_{eq}/\text{cm}^2$~\cite{Munker:2019vdo}. Both the NGAP and XDPW processes have a much faster charge collection with respect to the standard sensor.

\section{The MALTA2 chip}
\begin{figure}[htbp]
\centering 
\subfloat[Threshold]{\includegraphics[width=.4\textwidth]{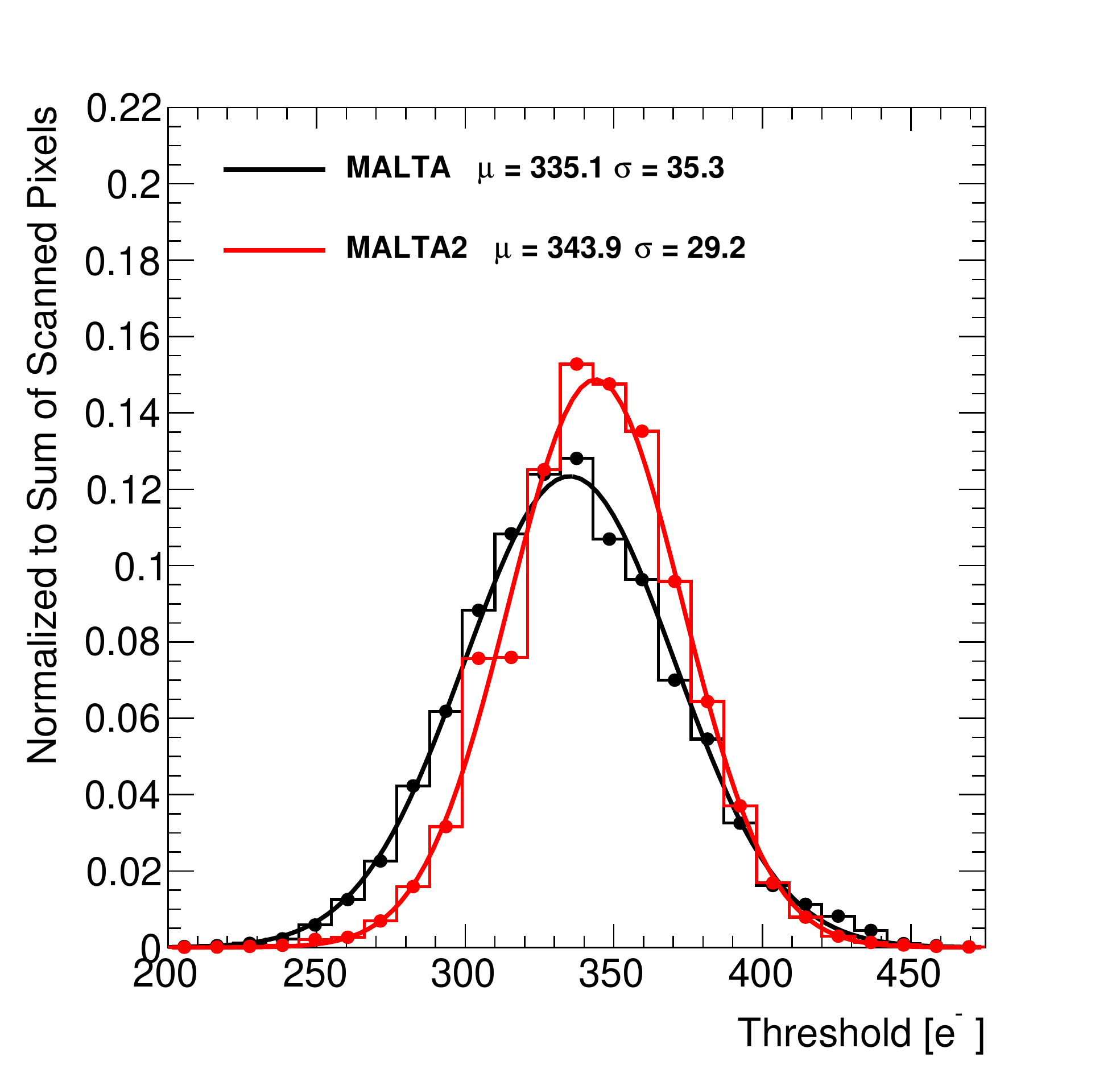}\label{fig:3a}}
\subfloat[Noise]{\includegraphics[width=.4\textwidth]{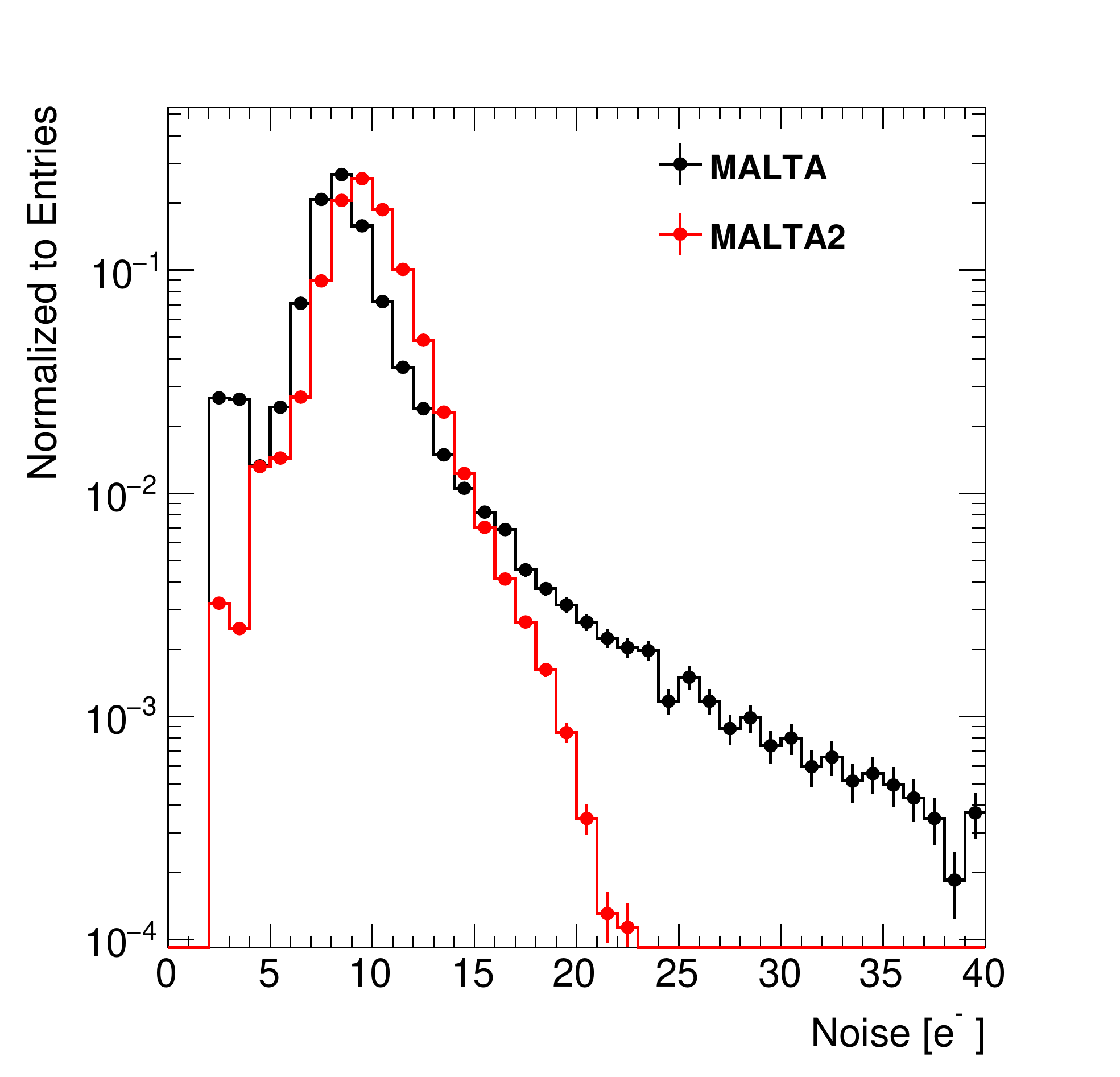}\label{fig:3b}}\\
\caption{Threshold (\ref{fig:3a}) and noise (\ref{fig:3b}) distributions of MALTA (black) and MALTA2 (red) samples. Both chips are un-irradiated, EPI, NGAP, 300~$\mu$m thick and operated the same setting values at the substrate voltage of $-6$~V. The RMS values of the noise distributions are about 3.5~e$^-$ and 2.25~e$^-$ for MALTA and MALTA2, respectively.}
\label{fig:3}
\end{figure}

The MALTA2 chip is the last generation of full-scale prototype produced. It integrates a matrix of $224 \times 512$ pixels in a total area of $10.12 \times20.2~\text{mm}^2$. An asynchronous read-out architecture is designed avoiding the propagation of the clock in the matrix to reduce the digital power consumption. As soon as a particle crosses the sensor, the in-pixel digital circuitry transmits the hit information from the chip to the periphery through a pattern of short pulses corresponding to the pixel address.
Pixels are organised in a dedicated group-logic allowing to operate at hit rates well above 100~$\text{MHz} / \text{cm}^2$.
The main difference with respect to the first generation of the MALTA chip prototype is in the front-end. It consists in the addition of a cascoded stage in the input branch and enlarged transistors in the feedback loop of the amplifier.
This allowed lower noise and enhanced gain, enabling the chip operation to thresholds down to $\mathcal{O}(100)~\text{e}^-$. Figure~\ref{fig:3} shows a comparison of the RTS noise and thresholds obtained with MALTA and MALTA2 chips. It was also demonstrated with an intermediate prototype, called mini-MALTA, that the enlarged transistors lead to a higher radiation tolerance~\cite{Dyndal:2019nxt}. With the same configuration and with chips irradiated at $10^{15}~\text{n}_{eq}/\text{cm}^2$ the standard transistor process reached an efficiency of about 87\% with a threshold of 340~e$^-$ whereas the new prototype provided an efficiency of about 98\% with a threshold of 200~e$^-$.

\section{Sample characterisation}

A campaign of test-beam measurements has been performed in 2021 and 2022 at the CERN Super Proton Synchrotron (SPS) exploiting the 180 GeV proton beamline to characterise the MALTA2 sensors in terms of their radiation tolerance and timing performance.
A custom pixel telescope consisting of six MALTA tracking planes (four Epi and two Cz samples) was used to study up to two MALTA2 devices under test at a time, hosted in a cold box. Behind the telescope layers a scintillator is installed. Located behind the telescope planes, the scintillator provides a timing reference for triggered signals.

\begin{figure}[htbp]
\centering 
\subfloat[Threshold]{\includegraphics[width=.4\textwidth]{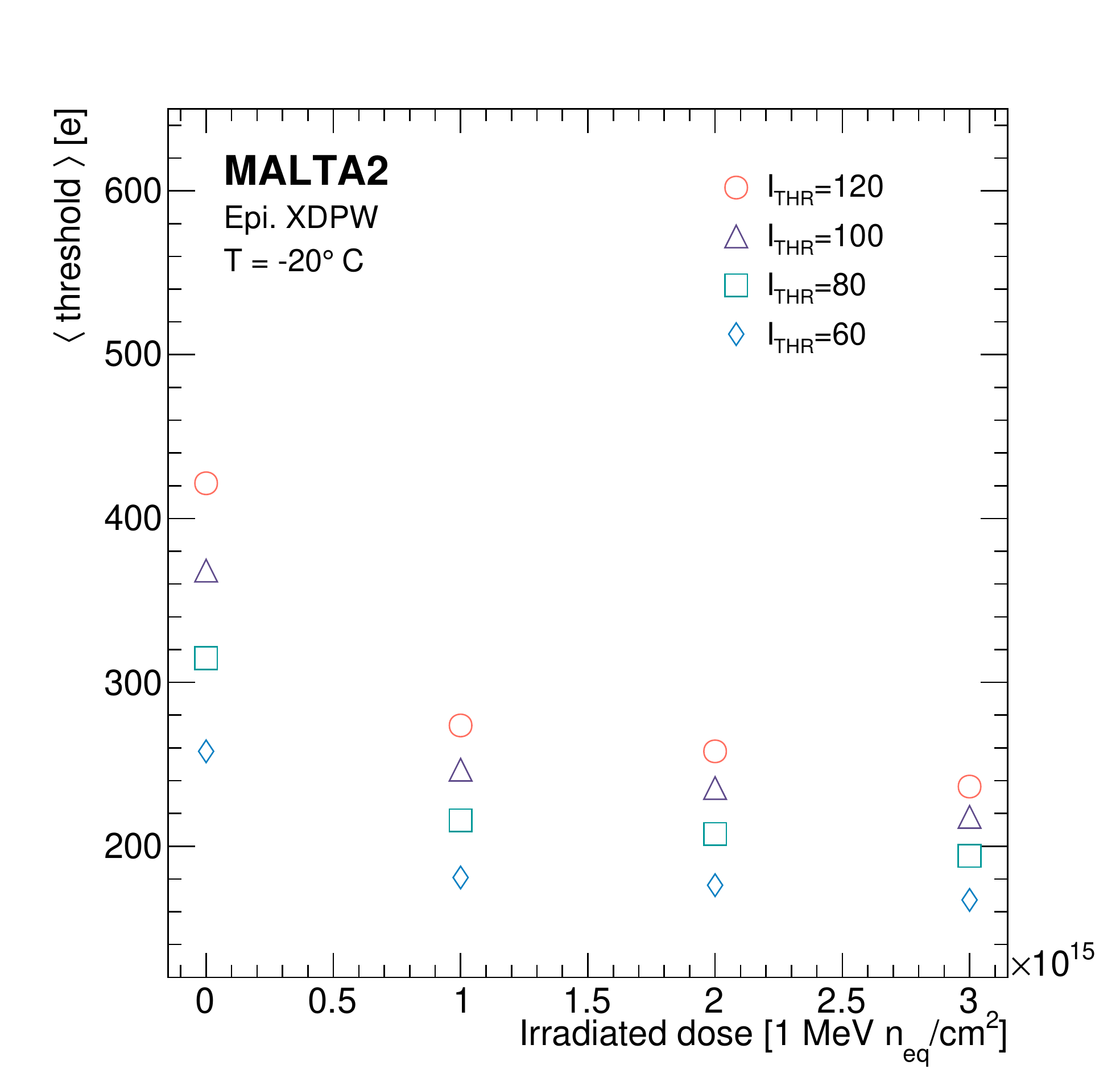}\label{fig:4a}}
\subfloat[Noise]{\includegraphics[width=.4\textwidth]{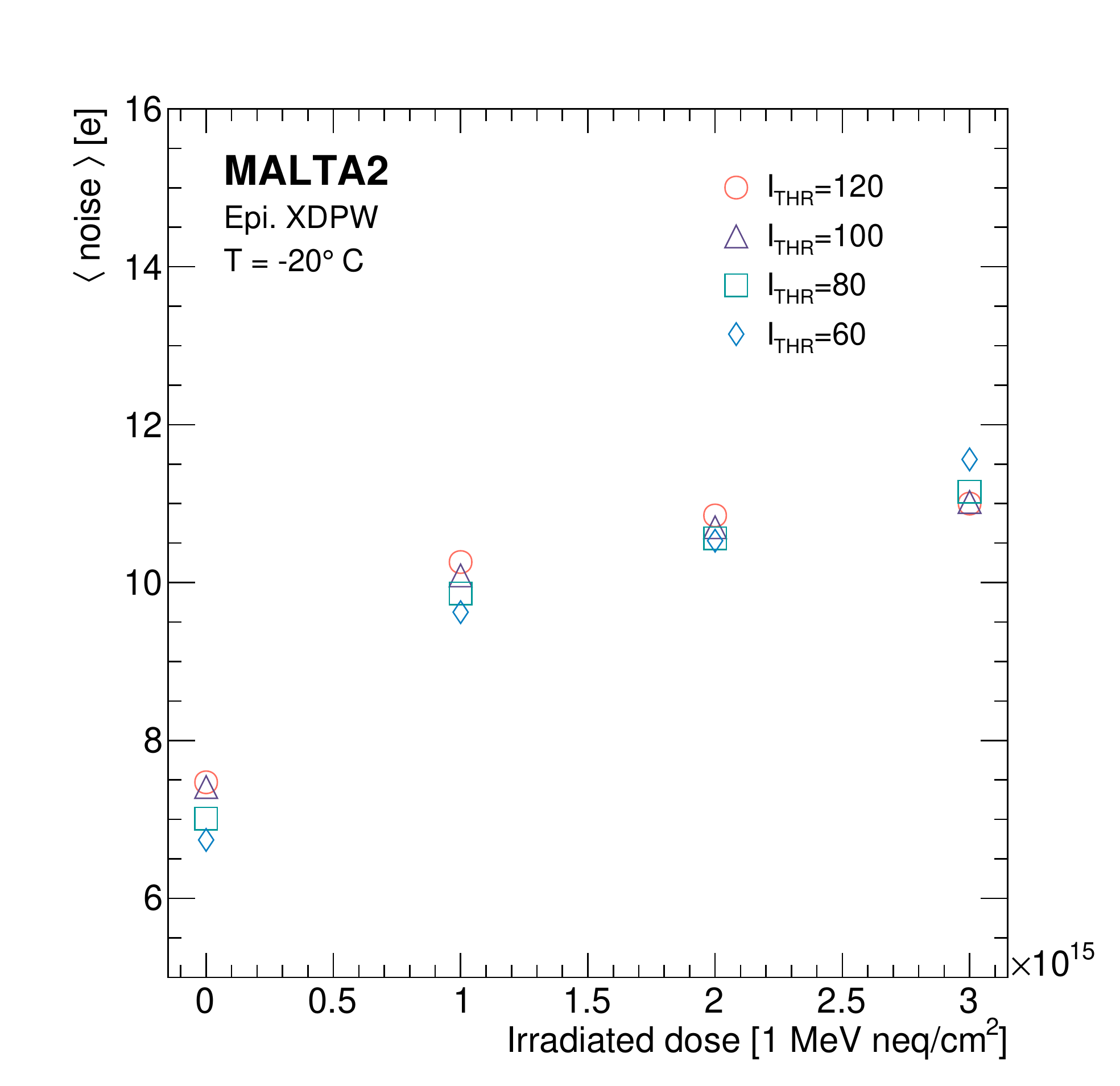}\label{fig:4b}}\\
\caption{ The average threshold (\ref{fig:4a}) and noise (\ref{fig:4b}) value of MALTA2 sensors on Epi substrate 100~$\mu$m thick with XDPW, as a function of the NIEL irradiated dose for several threshold settings at a substrate voltage of -6 V and a temperature of $-20\degree$C. For each $10^{15}~1~\text{MeV}~\text{n}_{eq}/\text{cm}^2$ the sensor receive 1~Mrad of gamma radiation.}
\label{fig:4}
\end{figure}

MALTA2 samples show a full efficiency on the entire matrix when un-irradiated already at a substrate voltage of $-6$~V for both Epi and Cz types.
The difference of the cluster size is instead evident between Cz and Epi samples. 
The Cz sample reaches a cluster size close to 2 for about 140~e$^-$ threshold and the Epi sample reaches a cluster size of 1.45 for 155~e$^-$ threshold. 
To check the impact of the irradiation on the samples, the threshold and noise measured are shown in figure~\ref{fig:4} as a function of the irradiated dose. The plots show that even at an irradiated dose of $3 \times 10^{15}~1~\text{MeV}~\text{n}_{eq}/\text{cm}^2$ and the 3~Mrad the threshold over noise ratio is greater than 10.

\section{Timing performance}

In order to measure the timing performance of the sensors a set of measurements are made to isolate the individual contributions.

\begin{figure}[htbp]
\centering 
{\includegraphics[width=.6\textwidth]{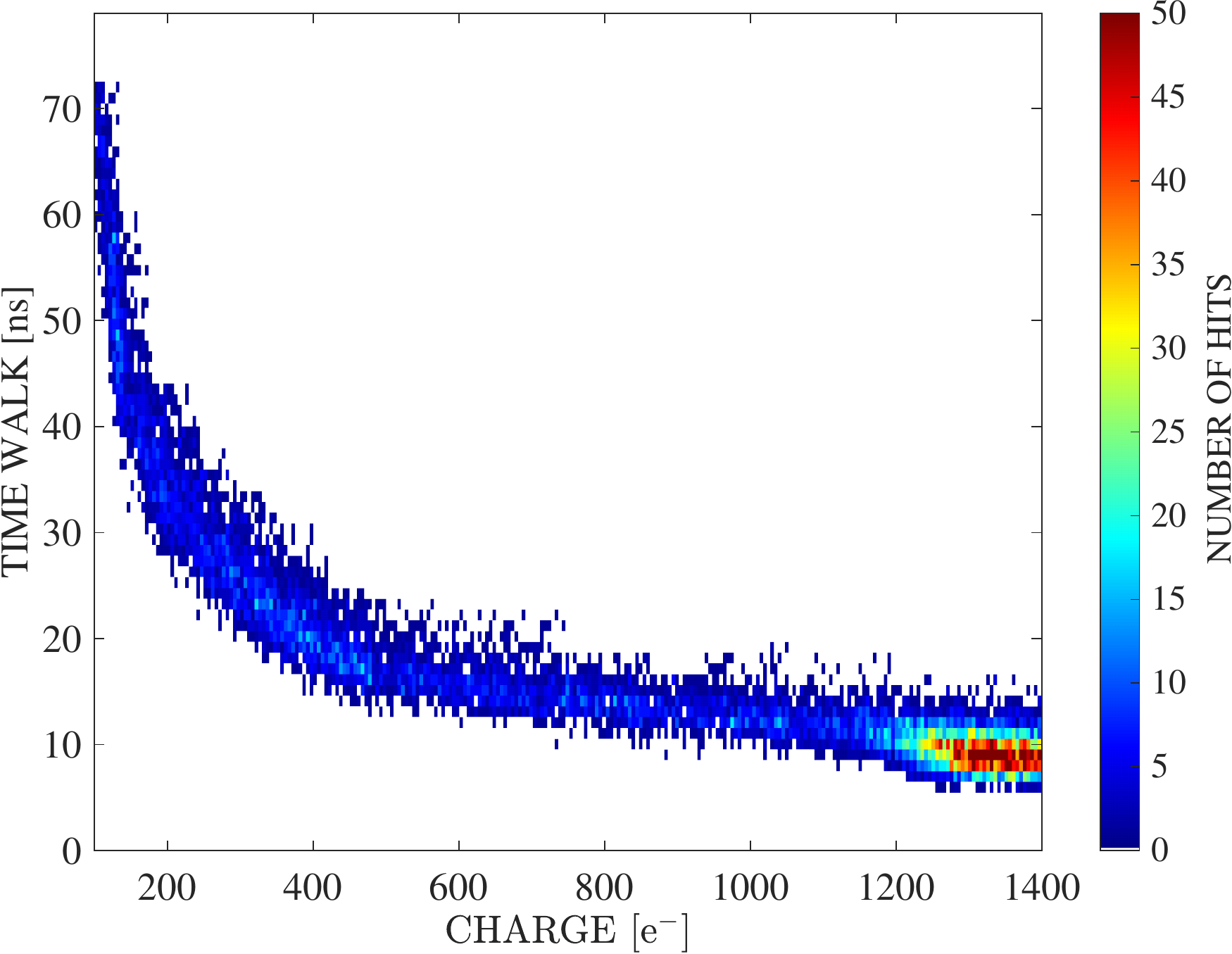}}
\caption{Time-walk as a function of the deposited charge obtained with a $^{90}$Sr source~\cite{Piro}.}
\label{fig:5}
\end{figure}

The time-walk of the front-end was measured using special pixels in the matrix with an analog output monitoring.
This shows how the time needed for the amplifier output to reach the discriminator threshold depends on the charge deposition (figure~\ref{fig:5})~\cite{Piro}.
A $^{90}$Sr radioactive source is used to produce MIP-like signals from the $\beta^-$ decay. About 16 thousands tracks were collected. The most probable value of charge deposition in the 30 $\mu$m thick sample used is around 1800~e$^-$ and the signal is collected by a cluster of up to four pixels.
Events with charges $\gtrsim 1200~\text{e}^{-}$ have a threshold crossing time of about 10 ns whereas about 90\% of the hits falls within a time window of 25 ns. Such events are referred to as in-time and the threshold corresponds to an input charge of about 200~e$^-$.
The residual 10\% is attributed to small charge deposition due to charge-sharing effects where smaller signals from pixel corners take more time to propagate.

The time jitter of the front-end electronics was estimated by injecting an increasing amount of charge within a pixel using the circuitry within the matrix digital read-out. 
The time of arrival of the generated hits from the injected charge is compared to the time reference of the charge injection trigger pulse, by using an external 3~ps binning TDC~\cite{Perktold:2014cya}.
A uniform response is observed across the entire chip.
The time jitter of the MALTA2 front-end electronics was measured to vary between 0.16 ns for injected charges above 1200 e$^-$ and 4.7 ns at a threshold of 100 e$^-$.

\begin{figure}[htbp]
\centering 
\subfloat[Epi, XDPW, 100 $\mu$m thick]{\includegraphics[width=.4\textwidth]{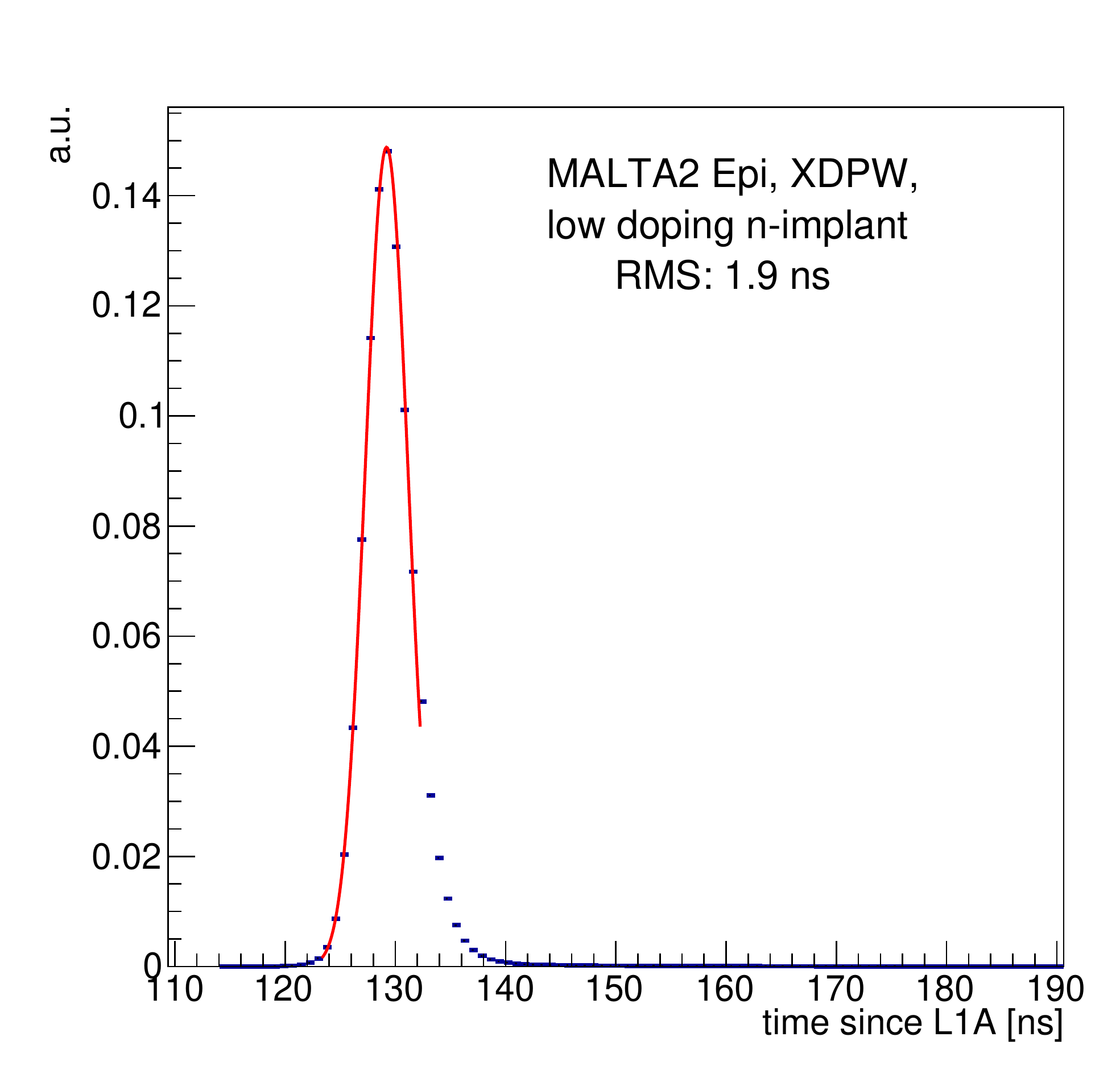}\label{fig:6a}}
\subfloat[Cz, XDPW, 100 $\mu$m thick]{\includegraphics[width=.4\textwidth]{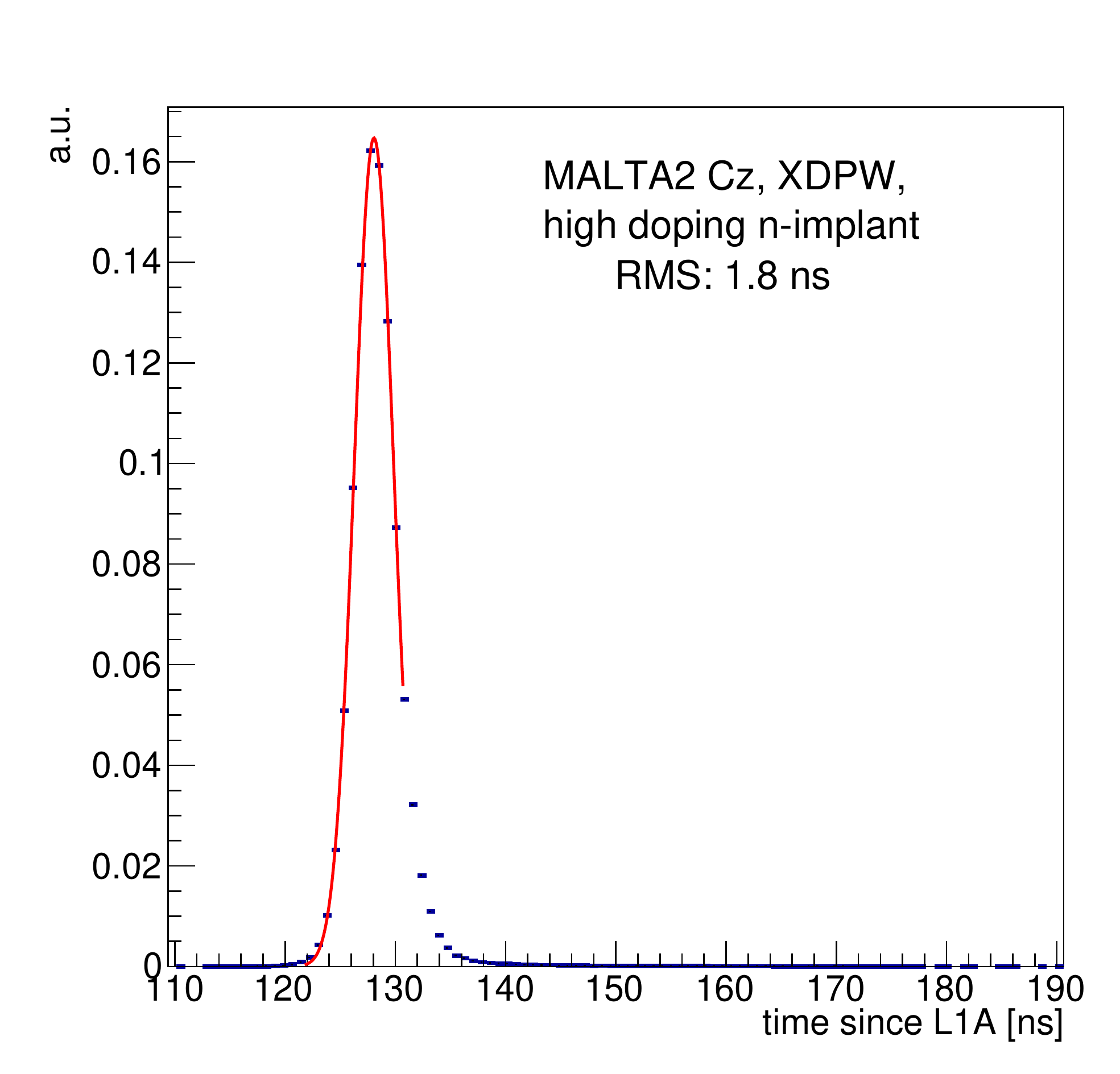}\label{fig:6b}}\\
\caption{Time of arrival of the leading hit in the cluster with respect to the scintillator reference. The quoted RMS value  is obtained by performing a Gaussian fit to the core of the distribution.}
\label{fig:6}
\end{figure}

\begin{figure}[htbp]
\centering 
\subfloat[Epi, XDPW, 100 $\mu$m thick]{\includegraphics[width=.4\textwidth]{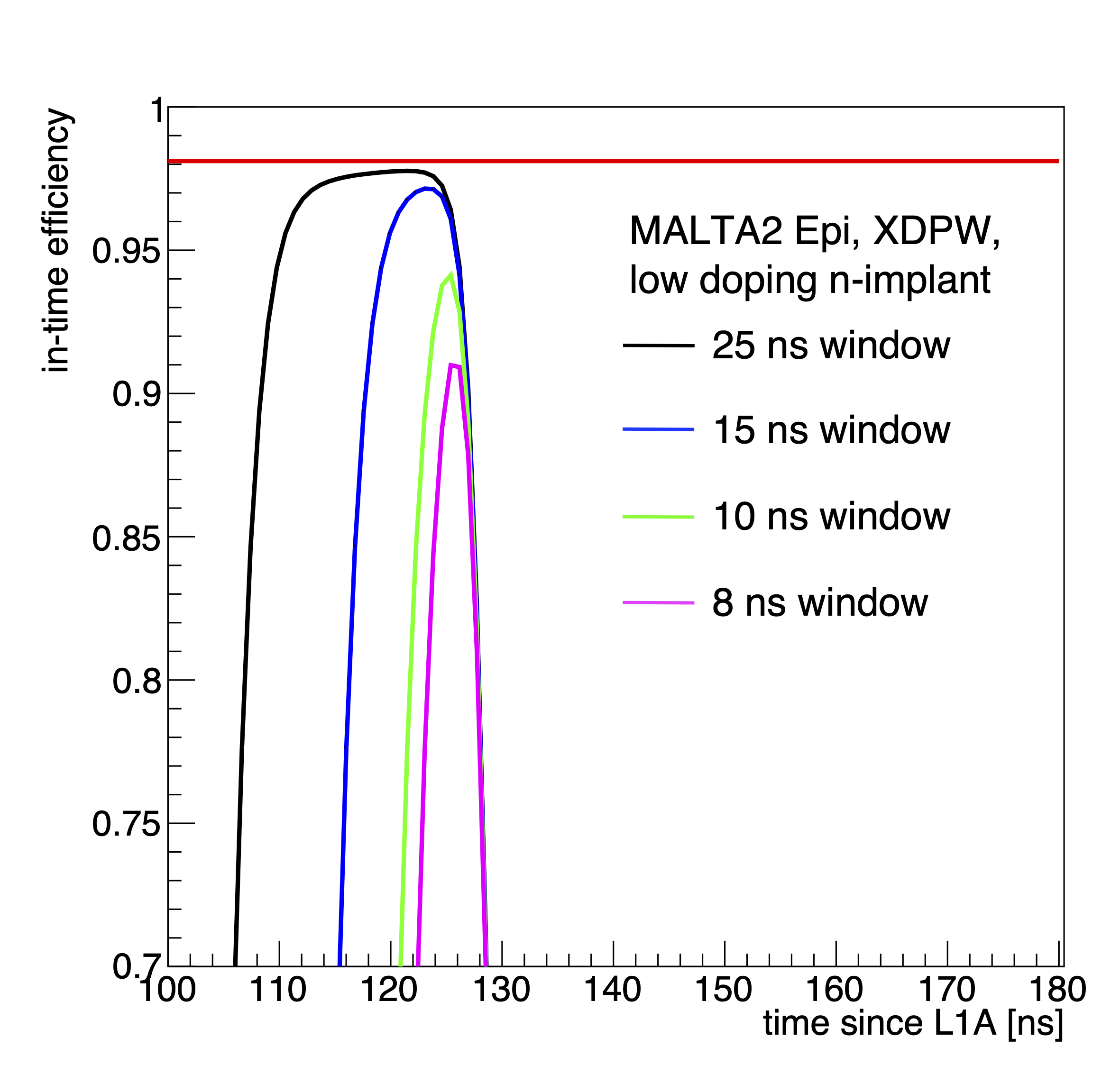}\label{fig:7a}}
\subfloat[Cz, XDPW, 100 $\mu$m thick]{\includegraphics[width=.4\textwidth]{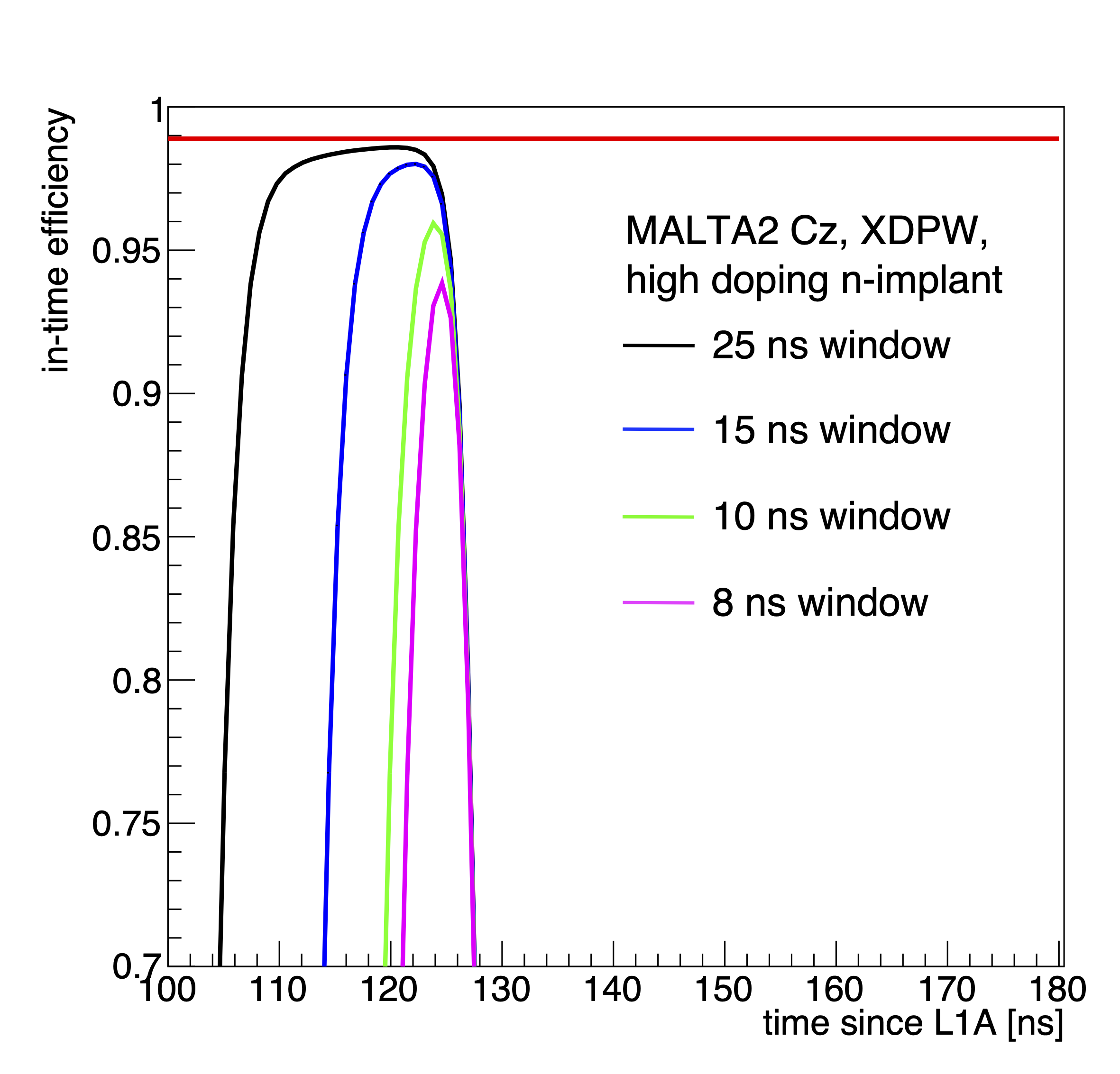}\label{fig:7b}}
\caption{In-time efficiency within a 25, 15, 10 and 8 ns windows in black, blue, green an pink lines, respectively.}
\label{fig:7}
\end{figure}

\begin{figure}[htbp]
\centering 
\subfloat[Epi, XDPW, 100 $\mu$m thick]{\includegraphics[width=.4\textwidth]{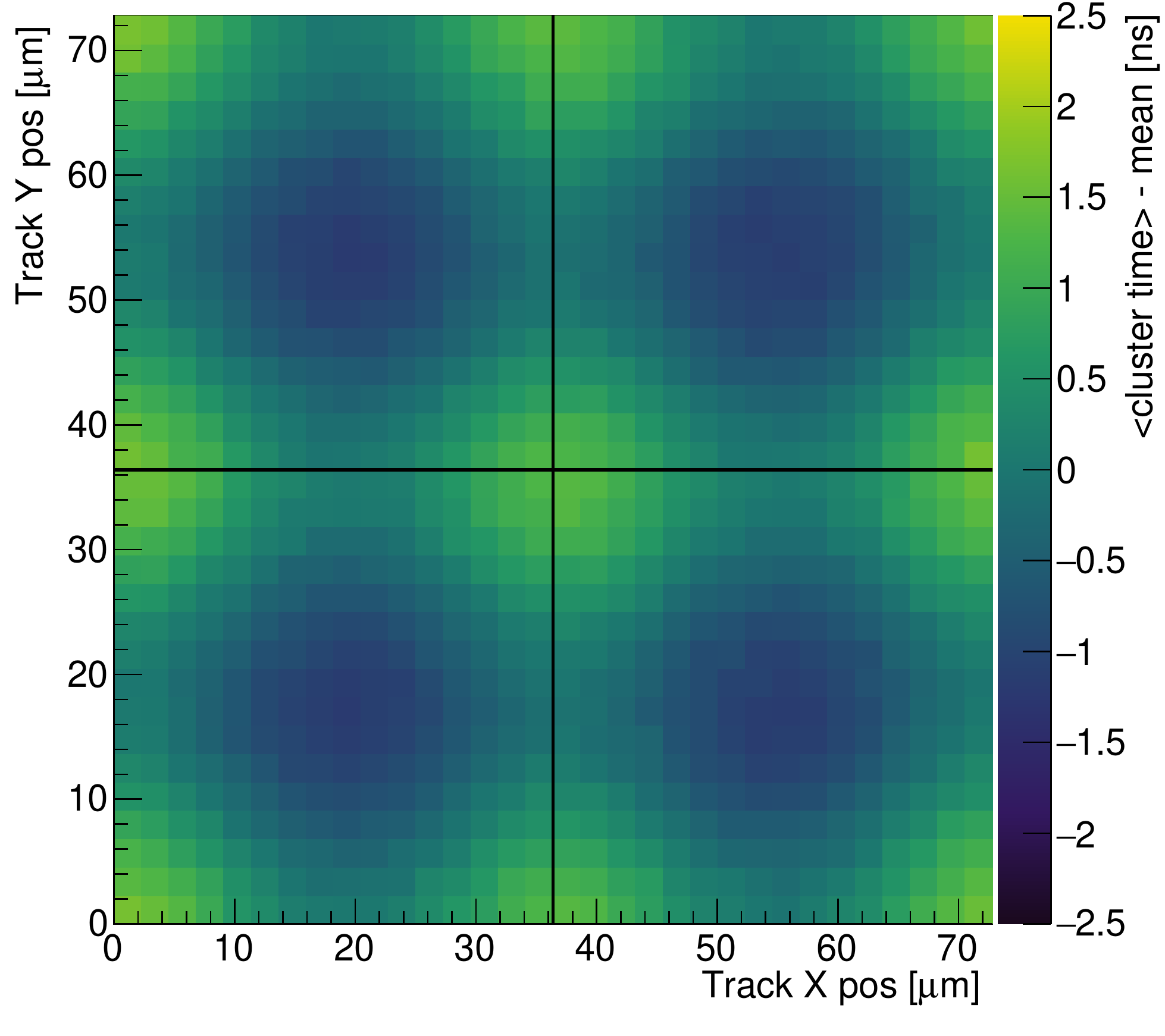}\label{fig:8b}}
\subfloat[Cz, XDPW, 100 $\mu$m thick]{\includegraphics[width=.4\textwidth]{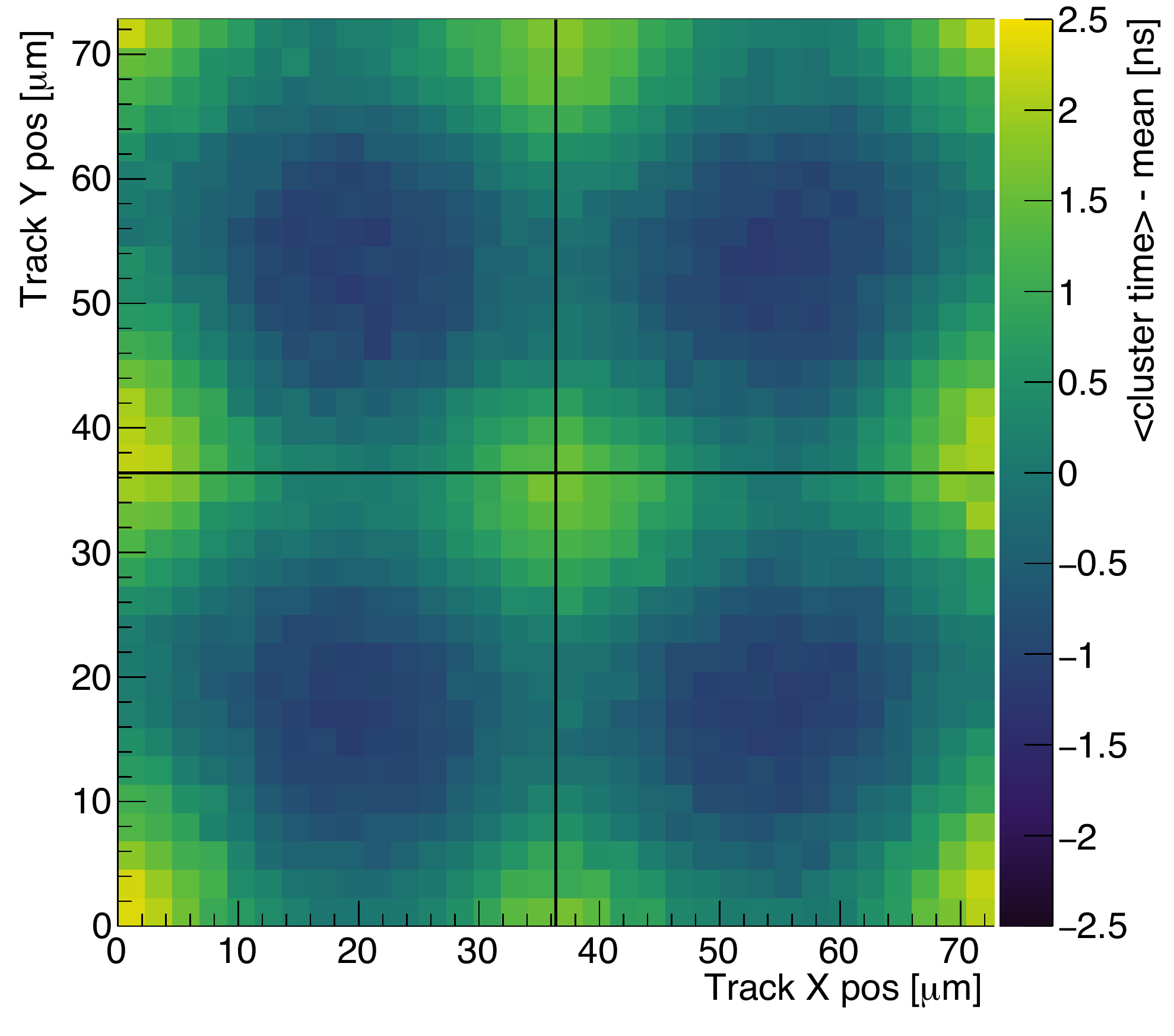}\label{fig:8a}}
\caption{Difference of timing of the leading hit in the cluster and the average timing over the entire chip projected in a $2\times2$ pixel matrix.}
\label{fig:8}
\end{figure}

Timing performance has also been measured during the test-beam campaign for un-irradiated chips. The time of arrival of the fastest hit in a pixel cluster with respect to the scintillator reference is shown in figure~\ref{fig:6} after having applied a correction that takes into account the time propagation of the hit information due to the structure of the chip read-out.
The performance is tested at a threshold value corresponding to 130~e$^-$ and 170~e$^-$ for the Epi and Cz chip, respectively.
The RMS of the time difference distribution  measured for the Epi and Cz MALTA2 samples is 1.9 ns and 1.8 ns, respectively. This RMS is the convolution of MALTA2 sensor intrinsic time-resolution including time-walk, charge collection time and electronics jitter. Sensor external effects like jitter of trigger scintillator (of about 0.5 ns) and sampling (of about 0.9 ns) jitters are further included in the shown RMS.
The in-time efficiency for both sensors is obtained by integrating the time of arrival distributions in different time windows.
As shown in figure~\ref{fig:7}, it is found to be greater than 98\% (90\%) for a 25 ns (8 ns) time window, suitable for applications at the LHC and other proposed future collider facilities.
Figure~\ref{fig:8} shows the projection over a $2\times2$ pixel matrix of the difference between the time of arrival of the leading hit in a pixel cluster and the average arrival time of signals over the entire chip.
A difference of 2-3 ns is observed between signals originated from the pixel centers and the ones from the corners. This is attributed to charge-sharing effect resulting in a lower charge deposition per pixel in the latter case leading to an increased time-walk.

\section{Conclusions}
MALTA2 is the latest full scale DMAPS prototype of the MALTA project.
A large set of characterisation studies is focused to estimate the performance of different pixel process modifications and substrate types, different chip configurations and NIEL radiation levels up to $3 \times 10^{15}~1~\text{MeV}~\text{n}_{eq}/\text{cm}^2$.
The time resolution is estimated to be below 2 ns.
The time-walk of the front-end electronics is found to be less than 25 ns for 90\% of the signals from a $^{90}$Sr source and the time jitter ranges between 0.16 and 4.7 ns for charge signals of 1200 and 100~e$^{-}$, respectively.
Many more results are in preparation from the full test-beam campaign at the 180 GeV proton beam at the CERN SPS.

\acknowledgments
This project has received funding from the European Union's Horizon 2020 Research and Innovation programme under Grant Agreement numbers 101004761 (AIDAinnova), 675587 (STREAM), and 654168 (IJS, Ljubljana, Slovenia).

\bibliographystyle{unsrt}
\bibliography{proceedings}
\end{document}